\documentclass[12pt]{article}
\usepackage{amsfonts}
\begin{document}
\begin{flushright}
FTUV-08/0218, IFIC-08-11   \\
Feb. 18, 2008\\
published in JHEP 04 (2008) 069\\
\end{flushright}

\vspace{3cm}

\begin{center}
{\Large\bf BPS Preons and the AdS-M-algebra} \\
 \vskip 1cm {\bf
Igor A. Bandos$^{\dagger\ast}$ and Jos\'e A. de
Azc\'arraga$^{\dagger}$ } \\
\vskip .5cm  {$^{\dagger}$ Departamento de F\'{\i}sica Te\'orica,
Univ.~de Valencia and IFIC (CSIC-UVEG), 46100-Burjassot
(Valencia), Spain
\\
$^{\ast}$Institute for Theoretical Physics, NSC Kharkov Institute
of Physics  and Technology,  UA61108, Kharkov, Ukraine}\\
bandos@ific.uv.es $\;,\;$ j.a.de.azcarraga@ific.uv.es \\

\vspace{3cm}

{\bf Abstract}

\end{center}
{\small

We present here the AdS generalization of BPS preons, which were
introduced as the hypothetical constituents of M-theory preserving
all but one supersymmetries. Our construction, suggested by the
relation of `lower dimensional preons' with higher spin theories,
can be considered as a deformation of the M-algebraic description
of the single supersymmetry broken by a preon, and provides
another reason to identify the AdS generalization of the
M-algebra, which we call the AdS-M-algebra, with $osp(1|32)$.

\vspace{1cm}

\newpage

\section{Introduction}

Preons were introduced \cite{BPS01} as the possible fundamental
constituents of M-theory. They are defined as BPS states that
preserve all supersymmetries but one. For $D$=11, this means $31$
supersymmetries out of $32$, and hence a preon may be labelled as
\begin{eqnarray}\label{Pr-def}
{} |BPS\; preon>  = |BPS\; , \; 31/32> \; . \qquad
\end{eqnarray}
As shown in \cite{BPS01}, a $k/32$-BPS state for $1<k<32$ may be
considered as a composite of ${\tilde n}=32-k$ preons. Fully
supersymmetric BPS states ($k=32$) do not contain any preons and,
hence, may be considered as {\it preonic vacua} (`vacua of vacua',
since all the $k$-supersymmetric BPS states are stable and are
considered themselves as different M-theory vacua); a preon is the
simplest excitation over such a fully supersymmetric vacuum. At
the other extreme, a non-supersymmetric (and, hence, non BPS)
state, breaking all 32 supersymmetries, is a composite of the
maximal number, $32$, of independent BPS-preons.

The preon definition \cite{BPS01}  also applies to arbitrary $D$
\cite{B02,30/32}. The $D$= 4,6,10 counterparts of a BPS preon can
be associated \cite{B02,30/32,BBdAST05} with an infinite tower of
free higher spin fields (see \cite{Vasiliev,Dima04}). This
identification can be established through  the quantization
\cite{BLS99,BBdAST05} of a generalized superparticle \cite{BL98}
which provides a model for a point-like or 0-brane preon
\cite{B02,30/32,BPS03}.

The standard realization of BPS states is provided by
$k$-supersymmetric solutions of the equations of motion for the
$D$=11 or type II $D$=10 supergravities, which are low energy
limits of M-theory\footnote{We will not consider here the N=1,
$D$=10 supergravity-SYM interacting systems describing the low
energy limits of the two heterotic strings and type I `corners' of
M-theory.}. A $k$-supersymmetric BPS state, or $k/32$-BPS state,
may be described by a supergravity solution preserving a fraction
$k/32$ of the supersymmetries. The $k$-supersymmetric {\it
bosonic} solutions are characterized by $k$ bosonic {\it Killing
spinors}, which obey the generalized Killing spinor equation
\begin{eqnarray}\label{KillingS}
{} {\cal D}\epsilon_{_I}{}^\alpha := D\epsilon_{_I}{}^\alpha -
\epsilon_{_I}{}^\beta \, t_\beta{}^\alpha := &
d\epsilon_{_I}{}^\alpha - {1\over 4} \epsilon_{_I}{}^\beta \,
\Gamma_{ab}{}_\beta{}^\alpha \omega^{ab} - \epsilon_{_I}{}^\beta
\, t_\beta{}^\alpha =0 \; , \quad I=1,\ldots , k\; .
\end{eqnarray}
In eq. (\ref{KillingS}), ${\cal D}=d-w = D-t$ is the generalized
covariant derivative involving the generalized connection
$w_\beta{}^\alpha=\omega_\beta{}^\alpha +t_\beta{}^\alpha$, where
$\omega_\beta{}^\alpha= {1\over 4}  \omega^{ab}
\Gamma_{ab}{}_\beta{}^\alpha$ is the spin connection and $
t_\beta{}^\alpha $ is the tensorial contribution constructed from
the fluxes (the field strengths of the gauge fields in the
supergravity multiplets). In $D$=11 supergravity \cite{CJS} this
tensorial contribution reads
\begin{eqnarray}\label{11D:t=}
 t_\beta{}^\alpha &=&  {i\over 18} e^a F_{ab_1b_2b_3}
\Gamma^{b_1b_2b_3}{}_\beta{}^\alpha +  {i\over 144} e^a
\Gamma_{ab_1b_2b_3b_4}{}_\beta{}^\alpha F^{b_1b_2b_3b_4} \; ,
\end{eqnarray}
where $F_4=dC_3={1\over 4} e^{c_4}\wedge \ldots \wedge e^{c_1}
F_{c_1\ldots c_4}$ is the field strength of the three-form gauge
field $C_3$. In $D$=11, eq. (\ref{KillingS}) is the only
restriction for Killing spinors, while in $D$=10 type II and lower
dimensional cases, they also have to satisfy an {\it algebraic}
equation, $ \epsilon_{_I}{}^\alpha{\cal M}_{\alpha\beta}=0$, where
the matrix ${\cal M}_{\alpha\beta}$ is constructed from the
scalars and the field strengths (fluxes) of the gauge fields  of
the corresponding supergravity multiplets. A hypothetical preonic
solution (for $D$=11 or IIA, IIB for $D$=10) would have $31$
Killing spinors, $\epsilon_I{}^\alpha$. Since there is only one
bosonic spinor $\lambda_\alpha$ orthogonal to all of them,
\begin{eqnarray}\label{ep-lam}
\epsilon_{_I}{}^\alpha \lambda_\alpha=0\; , \qquad I=1, \ldots ,
31\; , \qquad \alpha=1, \ldots , 32 \quad ,
\end{eqnarray}
a preonic supergravity solution may also be characterized by such
a {\it preonic spinor} $\lambda_\alpha$.

Algebraically ({\it i.e.}, from the structure of the M-theory
superalgebra or `M-algebra' \cite{M-alg}), any $k/32$ is allowed
for a BPS state \cite{BL98', GH99}. However, only (bosonic)
solutions for the following number of preserved supersymmetries
have been found at present
$$
k= 0,1,2,3,4,5,6,\;  8, \; 10 , \; 12, \; 14,\; 16,\;   18, \; 20,
\; 22, \; 24, \; 26 , \; 28 , \;\;\;\;\;\; 32\;
$$
(see {\it e.g.} \cite{DuffG2} for further discussion); the preonic
solution is conspicuously missing in this list.

   The interest on the possible existence of $31/32$-supersymmetric
or preonic solutions began around 2003 \cite{Duff03,Hull03,BPS03}.
Recently, a series of no-go results have been obtained for the
`free', `classical' $D$=11 and $D$=10 type II supergravities
\cite{GGPR06IIB,BdAV06,BGGPR06D11,FOF07} \footnote{A very recent
paper \cite{28/32} states that the maximal fraction ($\not=
32/32$) of supersymmetries preserved by a solution of the (again,
free and classical) type IIB supergravity is $28/32$.}. These
results were obtained by looking at the consistency equation for
the Killing spinors, $\epsilon{\cal R}=0$, where the generalized
curvature ${\cal R}$ is calculated using the (free, classical)
supergravity equations of motion. However, for supergravity with
($\alpha^\prime$) corrections the integrability condition and the
equations of motion will be modified (see \cite{alpha}), and a
full analysis remains to be done. As a result, the existence of
preonic solutions remains open when corrections are
present\footnote{Let us also note that the above no-go statements
have always been made for purely bosonic supersymmetric solutions
{\it i.e.}, for supergravity configurations with all fermionic
fields equal to zero, a restriction not implied by the preon
conjecture.} (see \cite{BdAV06} and \cite{BdA=FPhBPS07} for
further discussion). Moreover, even the possible absence of {\it
preonic solutions} in the presence of corrections or sources from
superbranes would not preclude the preon hypothesis, as such a
`preon conspiracy' would still allow us to consider all
supersymmetric BPS states as composites of preons (in the same way
as, by way of an analogy, quark confinement does not prevent the
existence of quarks). Although a dynamical mechanism to construct
$k/32$-BPS states out of $31/32$-preons is lacking, a further
study of the properties of preons may shed light in this
direction. With this in mind, we consider in this paper the
problem of AdS generalization of BPS preon. Not surprisingly, the
AdS preon will turn out to be related to the description of free
massless conformal AdS higher spin theories
\cite{MishaInt,Misha+2000} in the AdS-version of tensorial
superspaces given by the $OSp$ supergroup manifolds
\cite{BLPS99,Misha+,Dima+MP,Misha07}. In fact, a dynamical model
for our AdS preon is provided by the `preonic superparticle' on
$OSp(1|32)$, as discussed in Sec. 6.

\medskip

Let us go back to the idea of preons as elementary `excitations'
over a fully supersymmetric vacuum. The supergravity solutions
that describe fully supersymmetric BPS states include
\cite{FOF+GP}, besides the Minkowski vacuum of superPoincar\'e
symmetry, the $AdS_{(p+2)}\times S^{(D-p-2)}$ spaces,
$(D,p)=(11,2),(11,5),(10,3)$, and the $pp$-wave spaces which will
not be considered here. Thus, preons may correspond to the
simplest excitations over the Minkowski vacuum or over an
$AdS\times S$ vacuum. However, their original definition referred
to the M-algebra \cite{M-alg}, which is a generalization of the
superPoincar\'e algebra\footnote{To be more precise, this
generalization of the superPoincar\'e algebra is given by the
semidirect sum of the M-algebra \cite{M-alg} and $so(1,10)$
(alternatively, one may take $GL(32,\mathbb{R})$, the M-algebra
automorphism group \cite{GLautom}, when no decomposition in gamma
matrices is assumed), which can be shown to be an {\it expansion}
\cite{expansions} of the $osp(1|32)$ superalgebra. The
(32+528)-dimensional M-algebra itself, which is the maximal {\it
central extension} of the abelian $\{Q,Q\}=0$ superalgebra of the
$32$ fermionic generators (see \cite{JdeA00}), is a {\it
contraction} of $osp(1|32)$. Such a contraction is possible
because the M-algebra and $osp(1|32)$ have the same dimension.}.
Although the M-algebraic language is universal (as suggested {\it
i.e.} by the study of M-brane and D-brane systems), and thus the
preon notion is not restricted to considering excitations over the
Minkowski vacuum, it is natural to ask ourselves whether preons
can be defined in terms of a generalization of the AdS
superalgebra. This is tied to the AdS generalization of the
M-algebra, which we will call the {\it AdS-M-algebra}. Our
conclusion, which follows from the BPS preon generalization to be
presented here, is that the AdS-M-algebra is to be identified with
$osp(1|32)$, which in our preonic context appears as a {\it
deformation} of the M-algebra. The algebra $osp(1|32)$ as a
generalized AdS superalgebra in $D$=11 had been proposed in
\cite{AdSG, AdSFP, BvP00} (see also \cite{ Bars99, Ho97, AdSM+,
M-AdS} for other related superalgebras). The $osp(1|32)$ algebra
had already been singled out in the original $D$=11 supergravity
paper \cite{CJS}, and used as a basis for a discussion of the
gauge structure of $D$=11 supergravity \cite{D'A+F82,BAIPV04} as
well as in early discussions of general supersymmetry algebras
\cite{vHvP82}; its relevance in M(atrix)-theory had been put
forward in \cite{To97}.

\section{BPS preons, the preonic supermultiplet and the M-algebra}

An abstract BPS preonic state may be characterized by a single
bosonic preonic spinor $\lambda_\alpha$,
\begin{eqnarray}\label{BPS=l}
{} |BPS\,\;  preon>  = |\; \lambda  > \; , \qquad
\end{eqnarray}
which is orthogonal to the $31$ bosonic spinors
$\epsilon_{_I}{}^\alpha$, $\epsilon_{_I}{}^\alpha
\lambda_\alpha=0\,$, which determine the 31 supersymmetries
preserved by the preon,
\begin{eqnarray}
\label{eQl=0} {} \epsilon_{_I}{}^\alpha\, Q_\alpha |\;BPS \;preon>
= 0 \; , \qquad I=1, \ldots , 31\; , \quad
\end{eqnarray}
({\it cf.} eq. (\ref{ep-lam})). Due to the above orthogonality,
eq. (\ref{eQl=0}) implies that $Q_\alpha |\; \lambda
>\; \propto \lambda_\alpha$. This may be expressed as
\begin{eqnarray}\label{Ql=llf}
Q_\alpha |\; \lambda  > = \lambda_\alpha |\; \lambda ^f >\; ,
\qquad
\end{eqnarray}
where $|\; \lambda ^f >$ is a state with odd Grassmann parity
(assuming that  the original preonic state  $|\; \lambda  >$ is
 bosonic, as befits a state corresponding
to a purely bosonic solution of supergravity. The simplest preonic
supermultiplet contains only two states, $|\; \lambda>$ and $|\;
\lambda ^f>$,
\begin{eqnarray}
\label{Spreon}
|| \lambda^{super} >> := \left(\matrix{|\lambda\; > \cr |\lambda^f
> }\right)\,,  \qquad
\end{eqnarray}
with the action of the supersymmetry generator on $|\; \lambda^f>$
being defined in terms of the same bosonic spinor
$\lambda_\alpha$,
\begin{eqnarray}\label{Ql=ll}
{}  Q_\alpha |\; \lambda  >   = \lambda_\alpha |\; \lambda ^f > \;
, \qquad Q_\alpha |\; \lambda^f  >   = \lambda_\alpha |\; \lambda
> \; . \qquad
\end{eqnarray}
These supersymmetry transformations may be collected in one compact
equation
\begin{eqnarray}\label{QSpreon}
Q_\alpha ||\lambda^{super}>> = {\chi} {\lambda}_\alpha
||\lambda^{super} >>
 \; , \qquad {\chi}{\chi}=1 \qquad ,
\end{eqnarray}
in terms of the preonic supermultiplet $||\lambda^{super}>>$ and a
Clifford algebra variable ${\chi}$. When the preonic
supermultiplet is represented by a column vector, as in eq.
(\ref{Spreon}), ${\chi}$ is realized as the $\sigma^1$ Pauli
matrix, ${\chi}=\left(\matrix{0 & 1\cr 1 & 0}\right)$.

Now, assuming that $\lambda_\alpha$ is a $c$-number,
\begin{eqnarray}
\label{Ql=lQ}
Q_\beta {\lambda}_\alpha \, = \, {\lambda}_\alpha Q_\beta\,
 \; , \qquad
\end{eqnarray}
we conclude that the supersymmetry transformations generate the
M-algebra,
\begin{eqnarray}\label{Malg}
{} \{ Q_\alpha \, , \, Q_\beta \}  = P_{\alpha\beta} \; , \qquad
[\; P_{\alpha\beta} \, , \, Q_\gamma ]= 0 \; ,  \qquad  {} [\;
P_{\alpha\beta} \, , \, P_{\gamma\delta} ]= 0 \; . \qquad
\end{eqnarray}
Indeed, using  (\ref{Ql=lQ}) we find from  (\ref{Ql=ll}) that both
the BPS preon and its superpartner are eignestates of the
generalized momentum $P_{\alpha\beta}$ (here characterized as the
most general {\it r.h.s.} for the  $ \{ Q_\alpha \, , \, Q_\beta
\} $ anticommutator). The common eigenvalue matrix of $|\lambda> $
and $|\lambda^f >$ is given by the tensor product
$\lambda_\alpha\lambda_\beta$ of two copies of $\lambda$,
\begin{eqnarray}\label{PPr=lPr}
\cases{P_{\alpha\beta} |\lambda\;  > =
{\lambda}_\alpha{\lambda}_\beta |\lambda\; > \,
 \cr P_{\alpha\beta} |\lambda^f > = {\lambda}_\alpha{\lambda}_\beta |\lambda^f> \, }
\qquad \Leftrightarrow \qquad  P_{\alpha\beta} ||\lambda^{super}>>
= {\lambda}_\alpha{\lambda}_\beta ||\lambda^{super} >>
 \; .  \qquad
\end{eqnarray}
As the preonic spinor $\lambda_\alpha$  is a $c$-number (eq.
(\ref{Ql=lQ}) also implies
$P_{\alpha\beta}\lambda_\gamma=\lambda_\gamma P_{\alpha\beta}$),
one easily finds that on a preonic state or on the preonic
supermultiplet $[P,P]||\lambda >>=0$. This implies $[P,P]=0$ if we
do not allow for the presence of other generators, since the
possibility $[P,P]= cP$, allowed by Grassmann parity conservation,
is ruled out because $\lambda$ is nonvanishing and
$[P,P]|\lambda>=c\lambda\lambda |\lambda>= 0$ would require $c=0$.

\section{The AdS-M-algebra as suggested by AdS preons}
\label{f2}

 The previous discussion shows that  the original definition
of the BPS preon \cite{BPS01} reproduces the M-algebra
(\ref{Malg}), which generalizes the superPoincar\'e algebra by
involving the generalized momenta generator
$P_{\alpha\beta}=P_{\beta\alpha}$. This includes, in addition to
the standard momenta generator $P_m$ (through
$P\!\!\!\!/{}_{\alpha\beta}=P_m\Gamma^m_{\alpha\beta}$), a set of
tensorial central charges that reflect the existence of extended
objects in M-theory: they can be realized as topological charges
for various branes \cite{dAGIT89}. For instance, the
$SO(1,10)$-covariant decomposition $P_{\alpha\beta} =
\Gamma^m_{\alpha\beta}P_m +
i\Gamma^{a_1a_2}_{\alpha\beta}Z_{a_1a_2} + \Gamma^{a_1\ldots
a_5}_{\alpha\beta}Z_{a_1\ldots a_5}$, obtained by using the $D=11$
gamma matrices, includes the two- and five-index central charges
$Z_{a_1a_2}$ and $Z_{a_1\ldots a_5}$. Their spatial components,
$Z_{i_1i_2}$ and $Z_{i_1\ldots i_5}$, and those of their duals,
$Z^{i_1\ldots i_9}\propto \epsilon^{0i_1\ldots i_9j} Z_{0j} $ and
$Z^{i_1\ldots i_6}\propto \epsilon^{0i_1\ldots i_6j_1\ldots j_4}
Z_{0 j_1\ldots j_4}$, reflect, respectively, the existence of the
M2-brane (eleven-dimensional supermembrane), the M5-brane, the
Horava-Witten hyperplanes (M9-branes) and the Kaluza-Klein
monopole (KK6-brane) \cite{dAGIT89,Hull97,SorT}.

To look for the AdS generalization of the BPS preon we start from
the fact that, in lower dimensions $D$=4, 6 and 10, a BPS preon
wavefunction in its tensorial coordinate representation is given
by a scalar superfield on the corresponding tensorial superspace
$\Sigma^{({n(n+1)\over 2}|n)}$, $n$ =4, 8, 16, and can be
identified \cite{B02,30/32,BBdAST05} with a wavefunction
describing a tower of massless conformal higher spin fields
\cite{BLS99,BBdAST05} (see Sec. 5)\footnote{The case $n=2$
corresponds to a scalar superfield on $\Sigma^{(3|2)}$, which
coincides with the standard $D$=3 superspace, and no higher spin
fields appear.}. Now, the free AdS conformal massless fields can
be described in the same manner by the equations for a scalar
superfield on the $OSp(1|n)$ supermanifolds which, thus, provide
the AdS generalizations of the flat, tensorial
$\Sigma^{({n(n+1)\over 2}|n)}$ superspaces \cite{BLPS99,Misha+,
V01s,V01c,BPST04}. This suggests identifying an AdS preon state
with the one whose wavefunction is the $D$=11, $n$=32 counterpart
of the wavefunction describing, in lower $D$=4 and likely in
$D$=6,10 dimensions, towers of free conformal higher spin fields
in $AdS_{4,6,10}$ spacetimes respectively
\cite{Misha+,BPST04}\footnote{\label{hs} This is the case for
$D$=4, $n$=4. That a scalar field theory on the $OSp(1|n)$
supermanifold for $n=8,16$ describes the $D$=6,10 free massless
conformal AdS higher spin theories has still to be proven ({\it
e.g.}, by methods similar to those used in \cite{BBdAST05} to show
that a scalar field on the flat $n=8,16$ tensorial spaces
describes free conformal higher spin theories in $D$=6,10
Minkowski spaces, respectively).}.

The first consequence of this assumption is the identification of
the {\it AdS-M-algebra}. We conclude from the AdS preonic point of
view that the appropriate AdS generalization of the M-algebra (see
\cite{To97,AdSG,AdSFP,BvP00,AdSM+} for earlier discussions), the
{\it AdS-M-algebra}, is the orthosymplectic $osp(1|32)$ one,
\begin{eqnarray}
\label{M-AdS}
&& {} \{ Q_\alpha \, , \, Q_\beta \}  = M_{\alpha\beta} \; , \quad
[\; M_{\alpha\beta}
\, , \, Q_\gamma ]= {2\over R} C_{\gamma (\alpha} Q_{\beta )}\; , \quad \nonumber \\
\label{MM-AdS} && {} [\; M_{\alpha\beta} \, , \, M_{\gamma\delta} ]=
{2\over R} (C_{\gamma (\alpha} M_{\beta )\delta }+ C_{\delta
(\alpha} M_{\beta )\gamma })\; ,\qquad
\end{eqnarray}
where $C_{\alpha\beta}= - C_{\beta\alpha}$ is the nondegenerate
$32\times 32$ invariant $Sp(32)$ symplectic metric. The parameter
$R$ is introduced to make  the possibility of contracting
$osp(1|32)$ to the M-algebra (\ref{Malg}) \cite{To97} explicit. It
is convenient to take $R$ with dimensions of length; then it
corresponds to the radius of the generalized AdS space, for which
the $M_{\alpha\beta}$ play the r\^ole of isometry generators. In
the $R\rightarrow \infty$ limit the $M_{\alpha\beta}$ symplectic
generators of $osp(1|32)$  become the abelian generalized momenta
$P_{\alpha\beta}$. Reciprocally, $osp(1|32)$ is a deformation of
the M-algebra characterized by the radius deformation parameter
$R$. Algebra contractions abelianize part of the generators, and
deformations go in the inverse direction; in view of this, it is
not surprising that the AdS preon turns out to be a
non-commutative deformation of the original M-algebra preon
definition \cite{BPS01}. Let us note, to avoid confusion, that
this AdS preon does not correspond to a solution of some
`deformed' supergravity, but rather to a possible solution of
standard supergravity albeit with higher order corrections and/or
brane sources.

\section{AdS preons}

The discussion in Sec. 2 indicates that the AdS generalization of
the BPS preon notion will require dropping the commutativity
property of the preonic spinor since, by assuming eq.
(\ref{Ql=lQ}), we arrived at the M-algebra from the preonic
supermultiplet.

Further, since we want that in the $R\rightarrow \infty$ limit the
AdS preonic supermultiplet becomes the M-algebra one, we shall
assume that the AdS supersymmetry generators transform the AdS preon
and its superpartner among themselves in a way similar to
(\ref{Ql=llf}), where now a noncommuting but still Grassmann even
preonic spinor ${\Lambda}_\alpha$ replaces the $c$-number
$\lambda_\alpha$,
\begin{eqnarray}\label{AdSpreon}
Q_\alpha |\lambda> = {\Lambda}_\alpha |\lambda\; ^f>\; , \qquad
Q_\alpha |\lambda \; ^f> = {\Lambda}_\alpha |\lambda\;> \; , \qquad
[ {\Lambda}_\alpha \, , \, {\Lambda}_\beta]\not= 0\; . \qquad
\end{eqnarray}
 To have a suitable $R\rightarrow \infty$ limit, we conclude that
$[ {\Lambda}_\alpha \, , \, {\Lambda}_\beta]\propto {1\over R}$.
As the required coefficient is a dimensionless antisymmetric
spin-tensor, it is natural to identify it with $C_{\alpha\beta}$.
In such a way we find the following commutation relations for the
${\Lambda}_\alpha$ spinor operator entering (\ref{AdSpreon}),
\begin{equation}
\label{AdSLambda} {} [ {\Lambda}_\alpha \, , \, {\Lambda}_\beta] =
- {i\over 2R}C_{\alpha\beta} \quad ,
\end{equation}
which can be realized by
\begin{equation}
\label{Realization} {\Lambda}_\alpha = \lambda_\alpha - {i\over
4R}C_{\alpha\beta}\; {\partial\over
\partial \lambda^\beta}  \quad .
\end{equation}
Notice that the replacement $\lambda_\alpha \rightarrow
{\Lambda}_\alpha$ can be treated as passing to the Moyal star
product,
\begin{eqnarray}\label{star}
\lambda_\alpha \,\cdot\; \rightarrow {\Lambda}_\alpha \,\cdot \,=\,
\lambda_\alpha\, * \quad ,
\end{eqnarray}
see \cite{Misha+}. Eqs. (\ref{AdSLambda}), (\ref{star}) are a {\it
deformation} of the abelian
$[\lambda_\alpha\,,\,\lambda_\beta]=0$, and so eqs.
(\ref{AdSpreon}), (\ref{AdSLambda}) constitute a deformation of
(\ref{QSpreon}) resulting from the non-commutativity of
${\Lambda}_\alpha$. In the $R\rightarrow \infty$ limit of the
deformation parameter, ${\Lambda}_\alpha$ becomes the commutative
preonic spinor $\lambda_\alpha$ of the previous `flat' case. Thus,
the flat limit of the AdS preon reproduces the original
M-algebraic BPS preon definition \cite{BPS01}, of which the AdS
preon is a deformation.

 Denoting the AdS preonic supermultiplet also by $||\lambda^{super}>> $,
 as in eq. (\ref{Spreon}), the two equations in eq.
(\ref{AdSpreon}) are collected in a single equation ({\it cf.}
(\ref{QSpreon})),
\begin{eqnarray}\label{QSprAdS}
Q_\alpha ||\lambda^{super}>> = {\chi} {\Lambda}_\alpha ||\lambda
^{super} >>
 \; , \qquad {\chi}{\chi}=1 \; , \qquad {\Lambda}_\alpha =
 \lambda_\alpha - {i\over 4R}C_{\alpha\beta}\; {\partial\over
\partial \lambda^\beta}\quad , \qquad
\end{eqnarray}
which involves the  Clifford algebra element ${\chi}$ (see eq.
(\ref{QSpreon})) and the non-commutative preonic spinor
${\Lambda}_\alpha$. Thus, the AdS preonic supermultiplet is
associated with the following representation of the generators of
the $osp(1|32)$ algebra (eq. (\ref{MM-AdS}))
\begin{eqnarray}\label{OSP-pr}
Q_\alpha  = {\chi} {\Lambda}_\alpha\; , \quad M_{\alpha\beta}=
2{\Lambda}_{(\alpha}{\Lambda}_{\beta)} \quad  ,
\end{eqnarray}
where $\Lambda_\alpha$ satisfies the algebra of eq.
(\ref{AdSLambda}) and $\chi^2=1$. In matrix form, the preonic
realization of the $osp(1|32)$ generators reads
\begin{eqnarray}\label{OSP-prM}
Q_\alpha  = \left(\matrix{ 0 & {\Lambda}_\alpha \cr
{\Lambda}_\alpha & 0 }\right) \; , \qquad M_{\alpha\beta}=
\left(\matrix{2{\Lambda}_{(\alpha}{\Lambda}_{\beta)} & 0 \cr 0 &
2{\Lambda}_{(\alpha}{\Lambda}_{\beta)}}\right) \; . \qquad
\end{eqnarray}
The basic commutation relations of  ${\Lambda}_\alpha $ together
with the representation of $M_{\alpha\beta}$ in (\ref{OSP-pr}) are
collected in the multiplication table
\begin{eqnarray}\label{LL=}
{\Lambda}_{\alpha}{\Lambda}_{\beta}= - {i\over 4R}C_{\alpha\beta}
+ {1\over 2}M_{\alpha\beta}\; .
\end{eqnarray}

 \section{BPS preons, tensorial superspaces and
 massless conformal higher spin fields}

Our AdS generalization of the M-algebraic BPS preon, eq.
(\ref{AdSpreon}), and its associated AdS-M-algebra, are suggested
by the properties of higher spin theory as described by scalar
superfields in tensorial superspaces. This will be shown in this
section, which we begin by considering the realization of the
M-algebra preon as a scalar superfield in flat, tensorial
superspace before moving to the AdS case in Sec. 5.5.

\subsection{Preonic superwavefunction in tensorial
superspace $\Sigma^{({n(n+1)\over 2}|n)}$}

Tensorial superspaces $\Sigma^{({n(n+1)\over 2}|n)}$ are
parametrized by $n(n+1)/2$ even spin-tensor coordinates
$X^{\alpha\beta}$ and by $n$ odd, fermionic coordinates
$\theta^\alpha$ (see {\it e.g.} \cite{BL98,BL98',JdeA00, 30/32}),
\begin{eqnarray}\label{Sigma(n)}
\Sigma^{({n(n+1)\over 2}|n)} \; = \{ \; (X^{\alpha\beta},
\theta^\alpha ) \; \}\; , \qquad X^{\alpha\beta}=
X^{\beta\alpha}\; , \qquad \alpha = 1,\ldots, n \; .
\end{eqnarray}
In $D$=4,10 and 11 the minimal spinors have $n$=4, 16 and 32
components, and their associated even coordinates
$X^{\alpha\beta}$ have 10, 136 and 528 components respectively.
These include, besides those of the spacetime $D$-vector,
additional bosonic tensorial coordinates. Specifically,
\begin{eqnarray}\label{Sigma(4)}
& D=\; 4\; : & \quad \Sigma^{(10|4)}  =   \{ (x^m\, , \, y^{[mn]}
\, , \theta^\alpha ) \}\; , \quad X^{\alpha\beta}=
x^m\gamma_m^{\alpha\beta} + y^{mn}\gamma_{mn}^{\alpha\beta} \; ;
\\ \label{Sigma(16)}
&D=10\;  : & \quad \Sigma^{(136|16)}   =  \{ (x^m\, , \,
y^{[mnpqr]} \, , \theta^\alpha ) \}\; , \quad X^{\alpha\beta}=
x^m\tilde{\sigma}_m^{\alpha\beta} +
y^{mnpqr}\tilde{\sigma}_{mnpqr}^{\alpha\beta} \; ; \qquad
\\ \label{Sigma(32)}
&D=11\;  : & \quad \Sigma^{(528|32)}   =  \{ (x^m\, ,  \, y^{[mn]}
\, , \, y^{[mnpqr]} \, , \theta^\alpha ) \}\; , \quad \nonumber \\
&& {} \hspace{4cm}  X^{\alpha\beta}= x^m\Gamma_m^{\alpha\beta} +
y^{mn}\Gamma_{mn}^{\alpha\beta} +
y^{mnpqr}\Gamma_{mnpqr}^{\alpha\beta} \; . \qquad
\end{eqnarray}

The generalized momentum and the supersymmetry generators can be
realized as differential operators in $\Sigma^{({n(n+1)\over
2}|n)}$,
\begin{eqnarray}\label{Q=P=}
& P_{\alpha\beta}=-i\partial_{\alpha\beta}  \; , \qquad Q_\alpha = \partial_{\alpha} -
i\theta^\beta \partial_{\alpha\beta} \; , \quad \hbox{where} \quad
\partial_{\alpha\beta}:=
{\partial\quad \over\partial X^{\alpha\beta}} \; , \quad
\partial_{\alpha} := {\partial\;\;\over\partial \theta^{\alpha}}\;
\quad
\end{eqnarray}
(these give $\{ Q_\alpha , Q_\beta\}=2P_{\alpha\beta}$, but the
inclusion of the 2 here simplifies the coefficients below). The
$(X^{\alpha\beta}, \theta^\alpha)$ coordinates representation of
the BPS preonic supermultiplet $||\lambda ^{super}>>$ wavefunction
is
\begin{eqnarray}\label{PreonicWF}
\Phi_{(\lambda,\chi )} (X,\theta ) = << X , \theta ||
\lambda^{super}>> \quad ;
\end{eqnarray}
notice that the ${\chi}$ dependence of the {\it l.h.s.} comes from
$||\lambda ^{super} >>$, see eq. (\ref{QSprAdS}).

The defining relation (\ref{PPr=lPr}) implies that $\Phi$
satisfies the differential superwave equation
\begin{eqnarray}
\label{dPhi=llPhi} (\partial_{\alpha\beta} - i
\lambda_\alpha\lambda_\beta) \Phi_{(\lambda,\chi )} (X,\theta )=0
\quad .  \qquad
\end{eqnarray}
This {\it preonic equation} \cite{30/32} coincides with the
unfolded equations for higher spin fields \cite{V01s,Misha07}
formulated in tensorial space\footnote{In refs.
\cite{V01s,V01c,Misha07} the unfolded equations are written in the
form $\left({\partial\over\partial X^{\alpha\beta} }- i
{\partial\over \partial \mu^\alpha} {\partial\over \partial
\mu^\beta}\right) C(X, \mu)=0$, which is related to the preonic
equation (\ref{dPhi=llPhi}) by a Fourier transformation in the
auxiliary bosonic spinor variable $\lambda_\alpha$.}; it appeared
for the first time in the quantization \cite{BLS99} of the
generalized superparticle model \cite{BL98} on tensorial
superspace $\Sigma^{({n(n+1)\over 2}|n)}$.

\subsection{A model for a pointlike BPS preon in tensorial superspace
$\Sigma^{({n(n+1)\over 2}|n)}$ }

The action for a superparticle in $\Sigma^{({n(n+1)\over 2}|n)}$
with one auxiliary bosonic spinor reads \cite{BL98}
\begin{eqnarray}\label{S=BLac}
S= \int d\tau \lambda_\alpha \lambda_\beta
 (\dot{X}^{\alpha\beta} - i \dot{\theta}^{(\alpha}\,{\theta}^{\beta )})
 \quad , \quad \alpha=1,\dots,n \qquad .
\end{eqnarray}
It describes a 0-brane preon \cite{B02,30/32} or {\it preonic
superparticle} since its ground state preserves $(n-1)$ out of $n$
supersymmetries. The $\Sigma^{({n(n+1)\over 2}|n)}$ superspace
preonic wavefunction is obtained from the quantization of the
0-brane model (\ref{S=BLac}). To exhibit this schematically let us
note that eqs. (\ref{PPr=lPr}), (\ref{dPhi=llPhi}) look as the
quantum mechanical representation of the generalized
Cartan-Penrose relation
\begin{eqnarray}\label{PenroseEq}
P_{\alpha\beta}-\lambda_\alpha\lambda_\beta\approx 0\; ,  \qquad
\end{eqnarray}
which appears as a primary constraint for the canonical
generalized momentum for $X^{\alpha\beta}$. Actually, the
situation is slightly more complicated, because this constraint is
not first class, and its conversion to a first class constraint
requires the addition of a new variable. We will just state the
results and refer to \cite{BLS99} for details.

The quantization of the pointlike preon model (\ref{S=BLac})
produces a superwavefunction $\Upsilon$ that depends on
${X}^{\alpha\beta}$, $ {\theta}^{\alpha}$, $\lambda_\alpha$, {\it
and} on an additional Clifford algebra variable $\tilde{\chi}$,
$\tilde{\chi}^2=1$, which is introduced  in the process of
converting the fermionic second class constraint into a first
class one. The wavefunction $\Upsilon$ satisfies the wave equation
\cite{BLS99} which results from imposing the $32$ fermionic {\it
first class constraints} of the converted system,
\begin{eqnarray}\label{DY=lcY}
(D_\alpha  - \tilde{\chi}\lambda_\alpha) \Upsilon (X,\theta,\lambda ,\tilde{\chi}) =0\;
, \qquad & D_\alpha := \partial_{\alpha} + i \theta^\beta
\partial_{\alpha\beta} \; ,  \qquad \tilde{\chi}^2=1 \; ,  \qquad
\end{eqnarray}
where $D_\alpha$ is the covariant derivative in tensorial
superspace commuting with the supersymmetry generator $Q_\beta$ in
(\ref{Q=P=}). Thus, eq. (\ref{DY=lcY}) is supersymmetry invariant
provided that $\tilde{\chi}$ is inert under supersymmetry (as the
bosonic spinor variable $\lambda_\alpha$ is). The consistency
conditions for the quantum fermionic first class constraints
(\ref{DY=lcY}) give the bosonic first class constraint
\begin{eqnarray}\label{dY=llY}
(\partial_{\alpha\beta} - i \lambda_\alpha \lambda_\beta) \Upsilon
(X,\theta,\lambda ,\tilde{\chi}) =0 \; ,  \qquad
\end{eqnarray}
a clear counterpart of (\ref{dPhi=llPhi}).

Although (\ref{DY=lcY}) is similar to (\ref{QSpreon}), it includes
the supersymmetric covariant derivatives $D_\alpha$ rather than
$Q_\alpha$ in (\ref{Q=P=}). To solve this, let us now observe that
the shift  of a Clifford algebra variable $\chi$ by a nilpotent
one $\psi$, $\chi\rightarrow {\tilde \chi} = \chi - \psi$, is
still a Clifford element if the shift anticommutes with $\chi$,
$\{\chi,\psi\}=0$. In the present case, and with ${\tilde \chi}=
\chi - 2\theta\lambda$, we find
\begin{eqnarray}\label{chi2=1}
(\chi - 2\theta\lambda)^2=1 \qquad \Leftarrow  \qquad \cases{  \chi^2=1\; , \qquad
(\theta\lambda)^2=0 \; , \cr \qquad \{ \chi \; , \; \theta\lambda\}= 0 \; .}   \qquad
\end{eqnarray}
With this in mind it is easily seen that eq. (\ref{DY=lcY}) gives
the coordinate representation of the transformation rules
(\ref{QSpreon}), $(Q_\alpha  - \chi\lambda_\alpha) \Upsilon=0$, of
the preonic supermultiplet provided we identify
\begin{eqnarray}\label{Y=Phi}
\Upsilon (X,\theta,\lambda ,\tilde{\chi}) =\Phi_{(\lambda,\chi )}
(X,\theta ):=  << X , \theta || \lambda ^{super} >> \; , \qquad
\tilde{\chi}= {\chi} - 2\theta\lambda\quad .
\end{eqnarray}
For $n=4, 8, 16$ ($D=4,6,10$) the above wavefunction, with the
additional projection condition $\Upsilon (X,\theta,\lambda
,\tilde{\chi})= \Upsilon (X,\theta,-\lambda ,-\tilde{\chi})$  (see
\cite{BLS99,BPST04} for a discussion), describes a tower of
massless, conformal higher spin fields \cite{BLS99,BBdAST05}.

\subsection{D=4,6,10 massless conformal higher spin fields
from the preonic wavefunction on $\Sigma^{({n(n+1)\over 2}|n)}$
($n=4,8,16$) }

A Clifford superfield \cite{Dima88} is a function depending on
Clifford algebra variables, like our $\Upsilon (\tilde{\chi})$ in
(\ref{Y=Phi}) with $\; \tilde{\chi}\tilde{\chi}=1$. It is similar
to the familiar superfields in that its series decomposition in
the Cifford algebra arguments is finite. In the present case,
where $\Upsilon (\tilde{\chi})$ depends on only one Clifford
variable, the superfield contains only two (superfield)
components,
\begin{eqnarray}\label{Y=phi+}
\Upsilon (X,\theta,\lambda ,\tilde{\chi}) = \Phi^0
(X,\theta,\lambda )  + \tilde{\chi} \Phi^1 (X,\theta,\lambda )
\quad .
\end{eqnarray}
Eq. (\ref{DY=lcY}) implies that
\begin{eqnarray}\label{Dphi=}
 D_\alpha \Phi^0 = \lambda_\alpha \Phi^1 \quad , \quad D_\alpha \Phi^1
 = \lambda_\alpha \Phi^0 \quad .
\end{eqnarray}
These equations can be solved by expressing, say, $\Phi^1 $ in
terms of $\Phi^0 $, although to write such an expression in a
$GL(n)$-covariant manner one has to introduce a bosonic spinor
$u^\alpha$ `dual' to $\lambda_\alpha$ ({\it i.e.}
$u^\alpha\lambda_\alpha=1$):  $\Phi^1 = -i u^\alpha D_\alpha
\Phi^0 $. Now applying $D_\beta$ to the first equation in
(\ref{Dphi=}) and using the second one, we find the following
equation restricting only the $\Phi^0(X,\theta,\lambda )$
superfield (see \cite{BPST04})):
\begin{eqnarray}\label{DDphi=}
 (D_\alpha D_\beta - \lambda_\alpha\lambda_\beta)\Phi^0 =0 \; . \qquad
\end{eqnarray}
The symmetric part of (\ref{DDphi=}) gives the preonic equation
(\ref{dPhi=llPhi}), while the antisymmetric part reads
$D_{[\alpha} D_{\beta ]} \Phi^0 =0 $. This equation was proposed
in \cite{BPST04} as a superfield generalization of the Vasiliev
field equations \cite{V01s,Misha07} for the wavefunctions
describing the towers of all the bosonic and fermionic conformal
higher spin fields in $D=4$ tensorial space. Indeed, the same
equation is obeyed by the wavefunction integrated over the bosonic
spinor space $\phi(X,\theta)=\int d^n\lambda\,
\Phi^0(X,\theta,\lambda)$,
\begin{equation}
\label{DDPhi=0}
D_{[\alpha} D_{\beta ]} \phi(X,\theta)  =0     \; .
\end{equation}
Inserting the superfield expansion
\begin{eqnarray}\label{Phi=b+}
\phi(X,\theta)  = b(X)+ \theta^\alpha f_\alpha (X) + \sum\limits_{i=2}^n
\theta^{\alpha_1} \ldots \theta^{\alpha_i} \phi_{\alpha_1\ldots \alpha_i}(X)\; ,
\end{eqnarray}
in eq. (\ref{DDPhi=0}), one finds \cite{BPST04} that the higher
components of the $\phi(X,\theta)$ superfield vanish,
$\phi_{\alpha_1\ldots \alpha_i}(X)=0$ for $i\geq 2$, and that the
first two obey the bosonic and fermionic Vasiliev equations
\cite{V01s}
\begin{eqnarray}\label{ddb=0}
\partial_{\alpha [\beta}\partial_{\gamma ]\delta} b(X)=0 \; , \quad
\partial_{\alpha [\beta}f_{\gamma ]}(X)=0 \; ,  \quad \alpha, \beta, \gamma,
\delta  =1,\ldots , n\; . \quad
\end{eqnarray}
The proof that for $n=4$ these equations give a tower of all the
$D=4$ massless higher spin fields was given in \cite{V01s}. That
the $n=8$ and $n=16$ equations also describe a tower of {\it
conformal} massless fields in $D=6$ and $D=10$
 was shown in \cite{BBdAST05}, to which we refer the reader for
 details.

\subsection{Continuous spectrum of the D=11 preonic superparticle}

The situation for the $D$=11, $n$=32 M-theoretic case is less
clear. What makes it different from the previous $D=3,4,6$ and
$10$ cases is that in $D$=11 the vector $\lambda\Gamma_m\lambda$
is {\it not} lightlike, $(\lambda\Gamma_m\lambda)^2 \not= 0$,
which means that $P_mP^m\not= 0$ for the $D$=11 spacetime momentum
$P_m = \Gamma^{\alpha\beta}_m P_{\alpha\beta} \propto
\lambda\Gamma_m\lambda$. Moreover, $P_mP^m$ becomes an arbitrary
constant for the tensorial superspace $\Sigma^{(528|32)}$
pointlike preon model of eq. (\ref{S=BLac}) \cite{BL98, BL98'},
which is said to have a `dynamically generated mass' \cite{m-T}.
This property is tantamount to having a continuous mass spectrum.
Since this is typical of a composite system, we arrive at a
complementary description of a BPS preon: albeit fundamental, it
possesses a property associated with composite systems. This
situation is not new: the $D$=11 supermembrane (M$2$-brane) was
considered, as a fundamental object, as a $D$=11 counterpart of
the $D$=10 fundamental string and, at the same time, it was shown
to have a continuous spectrum, a property that was explained in
the Matrix model conjecture in which the M2 brane is considered as
a composite of D$0$-branes (N=2, $D$=10 massive superparticles).
Such a D$0$-brane picture is dual to the one in which the
M$2$-brane is considered to be fundamental. As for preons, we also
have that an {\it elementary} preon state has components in {\it
all} the tensorial charges associated with the 1/2 BPS branes,
which are themselves {\it composite} in the preonic picture. We
note, however, this latter property is also shared with the
$D$=4,6,10 dimensional counterparts of the M-algebraic BPS preon,
which nevertheless do not possess a continuous mass spectrum and
rather describe towers of massless conformal higher spin fields as
already discussed. The above dual aspect of the preon holds for
$D$=11, the M-theory dimension.

\bigskip

The mechanism to construct $k/32$-BPS states with $1<k<31$ from
the BPS preons is unknown, and one of the motivations to study
further the properties of BPS preons is to look for new insights
in this direction. It is natural to assume that the reduced
supersymmetry of a $k/32$-BPS state containing ${\tilde n}=32-k$
preons is the result of some kind of `interaction' among them. If
so, a possible description of such an interaction in $D=11$ should
be similar to a theory of interacting higher spin fields in the
lower $D=4,6,10$ dimensions. It is known that a selfconsistent
interaction of higher spin fields is possible in AdS but not in
Minkowski spacetime \cite{Vasiliev} (the interaction depends on
the inverse of the cosmological constant \cite{F+V,Vasiliev}).
Thus, the search for a selfconsistent interaction of an infinite
tower of higher spin fields begins by formulating the {\it free}
equations for these fields in AdS spacetime or an AdS superspace.

\subsection{Equations for $AdS_4$ conformal higher spin fields
on the $OSp(1|4)$ supergroup manifold}
\label{5.4}

 The AdS generalization of the free higher spin equations
in tensorial superspace, eq.(\ref{DY=lcY}), was obtained in
\cite{Misha+}. In our notation it reads \cite{BPST04}
\begin{eqnarray}
\label{AdSDY=lcY}
(\nabla_\alpha  - \tilde{\chi}{\Lambda}_\alpha)
\Upsilon (X,\theta,\lambda ,\tilde{\chi}) =0\; , \; &
{\Lambda}_\alpha = \lambda_\alpha - {1\over 4R}C_{\alpha\beta}\;
{\partial\over
\partial \lambda^\beta} \; ,  \quad \tilde{\chi}^2=1 \; , \quad \alpha=1,\dots,n
\;,\quad
\end{eqnarray}
where $\nabla_\alpha$ is defined by the decomposition of the
exterior derivative acting on the $OSp(1|n)$ manifold,
\begin{eqnarray}
\label{extdecomp} d= E^{\alpha\beta} \nabla_{\alpha\beta} +
E^{\alpha}\, \nabla_{\alpha}  \quad ,
\end{eqnarray}
in terms of the left-invariant Maurer-Cartan (MC) forms
$(E^{\alpha\beta}\, , \, E^{\alpha})$. These satisfy the
$osp(1|n)$ MC equations,
 \begin{eqnarray}
 \label{ospMC}
 dE^{\alpha\beta} +
 {1\over R}C_{\gamma\delta}E^{\alpha\gamma}\wedge E^{\beta\delta}+
 i E^{\alpha}\wedge E^{\beta} =0 \quad , \quad
 {\cal D} E^{\alpha}:=dE^{\alpha} +
 {1\over R}C_{\gamma\delta}E^{\alpha\gamma}\wedge
 E^{\delta}= 0 \; .
\end{eqnarray}
The above $\nabla_\alpha$ and $\nabla_{\alpha\beta}$ satisfy the
$osp(1|n)$ superalgebra,
\begin{eqnarray}\label{osp(1|n)}
& {} \{ \nabla_\alpha , \nabla_{\beta} \} = 2i
\nabla_{\alpha\beta} \quad , \quad [ \nabla_{\alpha \alpha^\prime}
, \nabla_{\beta} ] = {2i\over R}\,
C_{\beta (\alpha } \nabla_{\alpha^\prime )} \quad , \quad  \\
\label{sp(n)} & {} [ \nabla_{\alpha \beta} , \nabla_{\gamma
\delta} ] ={2i\over R} C_{\alpha (\gamma } \nabla_{\delta )
\beta}+ {2i\over R} C_{\beta (\gamma } \nabla_{\delta ) \alpha}
\quad . \quad
\end{eqnarray}

Decomposing the Clifford superfield $\Upsilon$ (eq.
(\ref{Y=phi+})), it is found that its second component can be
expressed in terms of the first one (as in (\ref{Dphi=}) for flat
tensorial superspace) and that its first component obeys
\cite{BPST04}
\begin{eqnarray}\label{DDphi=}
 \left(\nabla_\alpha \nabla_\beta +
 {\Lambda}_\beta{\Lambda}_\alpha\right)\Phi^0 =0 \; . \qquad
\end{eqnarray}
The symmetric $(\alpha\beta)$ part of this equation gives
\cite{Dima+MP} the AdS preonic equation that generalizes
(\ref{dY=llY}),
\begin{eqnarray}\label{AdSdY=LLY}
\left(\nabla_{\alpha\beta} - {i\over 2} \left({\Lambda}_\alpha
{\Lambda}_\beta + {\Lambda}_\beta{\Lambda}_\alpha \right)\right)
\Phi^0 =0 \; . \qquad
\end{eqnarray}
The antisymmetric $[\alpha\beta ]$ part of (\ref{DDphi=}) gives the
AdS generalization of equation (\ref{DDPhi=0}) for a scalar
superfield in flat tensorial superspace proposed in \cite{BPST04},
\begin{eqnarray}\label{DDPhi+R-1}
\left( \nabla_{[\alpha }\nabla_{\beta ]}+ {i\over
4R}C_{\alpha\beta} \right) \Phi^0=0\; , \quad
\end{eqnarray}

The set of Eqs. (\ref{AdSDY=lcY}) and (\ref{AdSdY=LLY}) is
equivalent to the following one-form differential equation proposed
in \cite{Misha+}
\begin{equation}\label{(d-w)P=}
(d-\hat{w}_0) {\hat \Upsilon}= 0\quad , \quad
\end{equation}
where $\hat{w}_0$ is given by $\hat{w}_0= E^{\alpha\beta}
M_{\alpha\beta}+ E^{\alpha}\, Q_{\alpha}$ with $M_{\alpha\beta}= 2
{\Lambda}_{(\alpha} {\Lambda}_{\beta)}\,$, $\, Q_{\alpha}=\chi
{\Lambda}_{\alpha}$ (eqs. (\ref{OSP-pr})) and ${\Lambda}_\alpha$
obeys the commutation relations (\ref{AdSLambda}). ${\hat
\Upsilon}$ depends on the $(X^{\alpha\beta},\theta^\alpha)$
variables of the $OSp$ supergroup manifold, as well as on ${\tilde
\chi}$ and the operator $\Lambda$, which is why ${\hat \Upsilon}$
(denoted $|\Phi>$) was called Fock module in \cite{Misha+}. Eq.
(\ref{(d-w)P=}) can also be written in the form \cite{Misha+}
\begin{equation}
\label{(d-w)*P=}
 (d-w_0*) {\Upsilon} = 0\; ,
\end{equation}
where ${w}_0= E^{\alpha\beta}  \lambda_{\alpha} \lambda_{\beta}+
E^{\alpha}\, \chi \lambda_{\alpha}$ is now used with the star
product of Eq. (\ref{star}). The selfconsistency equations for
(\ref{(d-w)*P=}), $dw_0=w_0\,*\,\wedge w_0$, give the $osp$ MC
equations (\ref{ospMC}). The same equation without star product,
$(d-w_0)\Upsilon = 0$, which leads through its selfconsistency
condition to the MC equations of the tensorial superspace algebra,
describes free higher spin fields in flat Minkowski spacetime.
Thus, the transition from the Minkowski higher spin field
equations in flat tensorial superspace to the equations on the
$OSp$ supergroup manifold describing the higher spin fields in
$AdS_4$ is given by a deformation which introduces
non-commutativity (see \cite{Misha+2000}).

\medskip

Summarizing, the AdS preon of Sec. 4 can be described by the
scalar field theory on the $OSp(1|32)$ supergroup manifold. This
is the $n=32$ ($D=11$) element of a family of scalar field
theories on the $OSp(1|n)$ manifolds, the $n=4$ representative of
which, $OSp(1|4)$, describes the higher spin theory on $AdS_4$. As
for the $n=8$ and $n=16$ cases, $OSp(1|8)$ and $OSp(1|16)$, they
are likely to describe the corresponding massless conformal higher
spin theories on $AdS_6$ and $AdS_{10}$ spaces (see footnote
\ref{hs}).

\bigskip

\section{The AdS preon as a BPS state. Preservation of all but one
AdS supersymmetries. }

The preonic spinors ${\Lambda}_\alpha$ of the AdS preon are
non-commuting (eqs. (\ref{AdSpreon}) and (\ref{QSprAdS})) and so
are $M_{\alpha\beta}$ in $osp(1|32)$ (\ref{MM-AdS}) that replace
the commutative $P_{\alpha\beta}$ of the M-algebra (\ref{Malg}).
As a result, the `momenta' sector of the $osp(1|n)$ superalgebra
does not allow for the M-algebraic analysis in \cite{BPS01} and it
is not obvious how to relate our AdS preon with the preservation
of a fraction of the supersymmetries, a typical property of a BPS
state.

To clarify this point, let us use the fact \cite{Dima+MP} that the
scalar superfield equations on the $OSp(1|n)$ supergroup manifold,
eqs. (\ref{AdSDY=lcY}), (\ref{AdSdY=LLY}), appear in the
quantization of the generalized superparticle on the $OSp(1|n)$
supermanifold\footnote{\label{f6} The $OSp(1|n)$ supergroup
manifold is `$GL(n)$ flat' \cite{Dima+MP} and this allows to
relate the AdS and the flat tangent superspace versions of the
generalized superparticle or preonic 0-brane model of \cite{BL98}
(Eqs. (\ref{S=BLSOSpac}), (\ref{S=BLac})). After a Penrose twistor
transform in both of them, they are described by the same action
in terms of a real $OSp$ supertwistor \cite{BL98} (fundamental
representation of $OSp(1|2n)$). Then, the quantization in momentum
space gives the same wavefunctions, and the specific AdS or
Minkowski spacetime wavefunctions are obtained by defining
appropriate measures for the Fourier transforms that lead to their
coordinate representation \cite{Dima+MP}. This is related to the
fact that both the flat tensorial superspace
$\Sigma^{({n(n+1)\over 2}|n)}$ and the $OSp(1|n)$ supergroup
manifolds can be identified with different cosets $OSp(1|2n)/ [\,
GL(n)\subset\!\!\!\!\!\! \times {\check
\Sigma}^{(\frac{n(n+1)}{2}|n)}]$ of the $OSp(1|2n)$ supergroup
with respect to differently chosen tensorial superspace subgroups
$\check{\Sigma}^{(\frac{(n(n+1))}{2} |n)}$ in $OSp(1|2n)$. }
\cite{BLPS99}
\begin{equation}
\label{S=BLSOSpac} S= \int d\tau \lambda_\alpha \lambda_\beta
\hat{E}_\tau^{\alpha\beta} \quad , \quad
\end{equation}
where $\hat{E}_\tau^{\alpha\beta} d\tau$ is the pullback to the
worldline of the $E^{\alpha\beta}$ Maurer-Cartan form on $OSp$
(see eq.(\ref{ospMC}), {\it cf.} \ref{S=BLac}). This superparticle
has the properties of an AdS preon: its ground states preserve all
the supersymmetries but one, as reflected by the $31$
$\kappa$-symmetries possessed by the $n=32$ version of the
$OSp(1|n)$ model of Eqs. (\ref{S=BLSOSpac}) (see \cite{B02, 30/32}
for further discussion in the M-algebraic language).

To clarify this point, consider first the case of a pointlike
M-algebra preon. The preonic 0-brane action in flat tensorial
superspace is given by eq. (\ref{S=BLac}). A preonic BPS state can
be associated with a purely bosonic solution of the equations of
motion that follow from this action. This is preserved by the
supersymmetries which keep the fermionic field equal to zero,
$\theta(\tau)=0$. The complete set of fermionic symmetries of the
action (\ref{S=BLac}) include global supersymmetry $\varepsilon$
and local fermionic $\kappa$-symmetry. In the present case of flat
tensorial superspace, a general fermionic transformation
$\delta=\delta_\varepsilon+ \delta_\kappa$ reads \cite{BL98}
\begin{eqnarray}
\label{vth=susy+k} \delta \theta^\alpha = \delta_\varepsilon
\theta^\alpha + \delta_\kappa \theta^\alpha := \varepsilon^\alpha
+ \kappa^I(\tau) \epsilon_I{}^\alpha (\tau) \; , \quad
\epsilon_I{}(\tau)^\alpha \lambda_\alpha(\tau) =0\; , \qquad
\matrix{ I=1,\ldots , 31\; , \quad \cr \alpha =1,\ldots , 32\;
,\quad }
\end{eqnarray}
where the $31$ bosonic spinors $\epsilon_I{}^\alpha(\tau)$ are
defined by the condition of being orthogonal to $\lambda_
\alpha(\tau)$. Then the supersymmetry which is preserved by the
purely bosonic, $\theta^\alpha(\tau)=0$ ground state solution is
characterized by
\begin{equation}
\label{susy=k}
\varepsilon^\alpha =- \kappa^I \epsilon_I{}^\alpha
\qquad \Leftarrow \qquad \delta
\theta^\alpha = 0 \; . \qquad
\end{equation}
This supersymmetry depends on the 31 parameters $\kappa^I$ of
local fermionic $\kappa$-symmetry, which become constant on the
solution. The fermionic spinor $\varepsilon^\alpha$ is constant
and so should be $\kappa^I \epsilon_I{}^\alpha$; the r\^ole of the
equations of motion in the supersymmetry preservation is seen at
this point. Indeed, the auxiliary bosonic spinor $\lambda_\alpha$
is constant on-shell, $\partial_\tau \lambda_\alpha=0$, so that
the $31$ bosonic spinors $\epsilon_I^{\alpha}$ orthogonal to it
can be chosen constant as well and so $\kappa^I$ is also constant
(see eq. (\ref{susy=k})). Thus, the constant $\varepsilon$ of the
preserved rigid supersymmetry $\varepsilon^\alpha $ is the sum of
products of the odd, constant $\kappa^I $ with the constant
bosonic spinors $\epsilon_I{}^\alpha $. The presence of $31$ free
fermionic parameters $\kappa^I $ in (\ref{susy=k}) allows us to
state that a bosonic solution of the equations following from the
action (\ref{S=BLac}) preserves $31$ target space supersymmetries.
This property allows us to identify \cite{B02,30/32} the ground
state of the 0-brane model (\ref{S=BLac}) with a pointlike BPS
preon, as it preserves $31$ out of the $32$ supersymmetries.
Further, this preservation of the target supersymmetries
(tensorial superspace or M-algebra supersymmetries) may be
formulated in abstract quantum mechanical terms for a BPS preon
state, as in eq. (\ref{eQl=0}), without reference to any specific
coordinates or momenta representation.

The situation is different in the AdS case. An AdS preon appears
as bosonic solution of the equations of motion that follow from
the action (\ref{S=BLSOSpac}) on the $OSp(1|32)$ ($OSp(1|n)$)
supergroup manifold. The superparticle lagrangian is now given in
terms of the bosonic MC forms of $OSp(1|n)$ (eqs.
(\ref{extdecomp}). The action is again invariant under the $31$-
($(n-1)$-) parametric $\kappa$-symmetry transformations
characterized by \cite{BLPS99}
\begin{eqnarray}
\label{kappa-curv} i_\kappa E^{\alpha\beta}:= \delta_\kappa Z^M
E_M^{\alpha\beta}(Z)=0\, , \quad i_\kappa E^{\alpha}\lambda_\alpha
:= \delta_\kappa Z^M E_M^{\alpha}\lambda_\alpha =0\; , \quad
Z^M:=(X^{\alpha\beta},\theta^\alpha)\; , \qquad
\end{eqnarray}
which can be described in terms of $31$ ($n-1$) bosonic spinors
$\epsilon_I{}^\alpha$ orthogonal to $\lambda_\alpha$ as in eq.
(\ref{ep-lam}),
\begin{equation}
\label{kappa-eps}
 i_\kappa E^{\alpha}:= \delta_\kappa Z^M E_M^{\alpha}
 = \kappa^I (\tau)
\epsilon_I{}^\alpha(\tau)\; . \quad
\end{equation}
 The equations of motion for the action (\ref{S=BLSOSpac})
 include ${\cal D}(\lambda_\alpha\lambda_\beta)=0$, with the $Sp(n)$
 covariant derivative ${\cal D}\lambda_
\alpha:=d\lambda_ \alpha+ {1\over R} E_\alpha^{\cdot\,
\beta}\lambda_\beta$ as in (\ref{ospMC}). Since $\epsilon^\alpha_I
\lambda_\alpha=0$ , the bosonic `Killing' spinors
$\epsilon_I{}^\alpha$ in (\ref{kappa-eps}) are now covariantly
constant rather than constant, ${\cal D}\epsilon_I{}^\alpha
 :=d\epsilon_I{}^\alpha -
 {1\over R}\epsilon_I{}^\beta E_\beta^{\cdot\, \alpha}=0$
and hence they are $\tau$-dependent.

The transformation of the fermionic coordinate functions under the
$OSp(1|32)$ symmetry of the action reads $\delta_\varepsilon
{\theta}^\alpha (\tau)= \varepsilon^\alpha - {1\over R}
\varepsilon^\beta X_\beta{}^\alpha (\tau)  + O( {1\over R^2}) + O
(\theta\theta) $. Thus, the supersymmetries preserved by the
ground state with $\theta^\alpha =0$ are characterized by ({\it
cf.} eq. (\ref{susy=k}))
\begin{eqnarray}\label{e-cons=}
 \varepsilon^\alpha  =
 -\kappa^I \epsilon_I{}^\beta \left(\delta_\beta{}^\alpha +
  {1\over R} X_\beta{}^\alpha +
   O ({1\over R^2}) \right)\quad \Leftarrow
  \quad \delta_\varepsilon \theta^\alpha\vert_{\theta=0}=0 \; . \qquad
\end{eqnarray}
For finite $R$ the terms involving the explicit $X^{\alpha\beta}$
(and $\theta^\alpha$) dependence hamper the abstract quantum
mechanical description of the supersymmetries preserved by the AdS
preonic superparticle ground state. When $R\mapsto \infty$, in
which limit $\varepsilon^\alpha$ becomes constant, eq.
(\ref{e-cons=}) reproduces (\ref{susy=k}) and an abstract quantum
mechanical description of the preserved symmetries becomes
possible.

 Hence, our AdS preon {\it is} a BPS state preserving
31 (($n$-1) in general) supersymmetries. This can be seen in the
generalized coordinates representation of the preonic
superwavefunction or through the corresponding pointlike model of
eq. (\ref{S=BLSOSpac}), where one also observes (eq.
(\ref{e-cons=})) that the preserved supersymmetries are $X$- (and
$\theta$-)dependent. This shows why in the AdS case it is
difficult to describe the preserved supersymmetries in an abstract
quantum mechanical state terms. In other words, the above
discussion explains why representation of the $OSp$ supersymmetry
generators on the states in (\ref{AdSpreon}), which emphasize the
single broken supersymmetry, cannot be reformulated through the 31
preserved supersymmetries. Such a representation is provided,
instead, by a deformation of the M-algebraic definition of the
single supersymmetry broken by the BPS preon. This is obtained by
replacing the bosonic spinor $\lambda_\alpha$ by the
non-commutative preonic spinor $\Lambda_\alpha$ (eqs.
(\ref{AdSpreon}), (\ref{AdSLambda})) or by moving to the Moyal
product, $\lambda_\alpha \cdot \mapsto
\Lambda_\alpha=\lambda_\alpha *$, eq. (\ref{star}).

\section{Conclusions and discussion}

We have given here the AdS generalization of the M-algebraic
definition of the BPS preon. Although the M-algebra language is
meant to be universal (as suggested by the study of the 1/2-BPS
superbrane states), and so is the preon concept \cite{BPS01}, the
question of its AdS generalization arises naturally when
considering a preon as an excitation over a fully supersymmetric
AdS-type (rather than Minkowski) vacuum. We have then found that
the AdS preon is a deformation of the M-algebra one \cite{BPS01}
(as {\it e.g}, eq. (\ref{QSprAdS}) is a deformation of
(\ref{QSpreon})). This deformation character is exhibited by the
explicit presence of $1/R$ in all the AdS equations, which
reproduce those of the flat case in the $R\rightarrow \infty$
limit. Conversely, all our AdS equations are obtained by the
replacement of the $\cdot$ product by the star $*$ one, eq.
(\ref{star}), in the M-algebraic flat ones.

Our generalization is suggested by the observation that the
$D$=4,6,10 tensorial superspace counterparts of the M-algebra BPS
preon can be identified \cite{B02,30/32,BPST04} with the towers of
all the free massless, conformal higher spin fields in the
respective flat Minkowski spaces. In other words, the
wavefunctions of the $n=4$ and $n=8,16$ counterparts of the
M-algebra BPS preons in {\it flat} tensorial superspaces (the
manifolds of the rigid $\Sigma^{(\frac{n(n+1)}{2}|n)}$ tensorial
superspace groups) describe infinite towers of free conformal
higher spin field strengths in $D=4$ \cite{BLS99} (see also
\cite{V01s}) and in $D=6,10$ \cite{BBdAST05}. Similarly, we
identify the wavefunction of an AdS preon state with the
$OSp(1|32)$ counterpart of the scalar superfield on the $OSp(1|4)$
supergroup manifold which describes \cite{Misha+,Dima+MP,BPST04}
all the conformal higher spin fields in $AdS_4$ space. Thus, as
the generalized AdS geometry of the free $AdS_4$ higher spin
fields is described by the $OSp(1|4)$ supergroup manifold (and,
likely, the scalar superfield on $OSp(1|n)$ for $n=8,16$ describes
the AdS massless conformal higher spin fields in $D$=6,10 as
well), our construction indicates that the {\it AdS-M-algebra} is
given by $osp(1|32)$, in agreement with \cite{To97, AdSG, AdSFP}
(see also \cite{BvP00, AdSM+}).

   To see how to relate the AdS preon definition with the
preservation of a fraction of the supersymmetries, we have
discussed in Sec. 6 the superparticle model on the $OSp(1|n)$
supergroup manifold \cite{BLPS99,Dima+MP}. The ground state of
this model preserves $31$ supersymmetries associated with the
31-parametric $\kappa$-symmetry of its action. Therefore, it is a
BPS preon and the $OSp$ superparticle can be called an AdS preonic
0-brane. However, the action of this preserved part of the AdS
supersymmetry on this BPS preonic state is $X$- (and $\theta$-)
dependent, as it is the AdS supersymmetry acting on $OSp(1|n)$
supermanifold. Thus, it is hard to see this preserved
supersymemtry in the abstract (bra-ket) quantum mechanical
language used to define the AdS preon (although there is no
problem to describe it by a superwavefunction in the generalized
coordinate representation). This explains why the preonic
representation of the $osp$ supersymmetry generators
(\ref{OSP-pr}) cannot be obviously translated in terms of
preserved supersymmetries and leads instead to a non-commutative
{\it deformation} of the M-algebraic definition of the BPS preon,
singling out the supersymmetry {\it broken} by the AdS preon. The
appearance of a deformation is again not surprising if we recall
that the Moyal brackets were introduced in higher spin theory
\cite{Misha+2000} to describe the free $D$=4 higher spin theories
in $AdS_4$ space.

The notion of the AdS preon introduced here suggests that the
search for a dynamical mechanism to obtain the $k/32$-BPS states
from the BPS preons may be related to the problem of constructing
a consistent interaction theory of a tower of  massless conformal
higher spin fields. Interacting, massless conformal higher spin
theories were constructed in \cite{MishaInt}. However, in our
preonic context, we need a formulation of such an {\it
interacting} theories in tensorial superspaces (see
\cite{BLS99,V01s,V01c,BPST04,BBdAST05} for the free case). This is
still unknown, although progress in this direction has been made
by introducing higher spin gauge potentials in generalized AdS
superspace \cite{Misha07}.

A natural development of the present work would be to look for
composites of AdS preons, in particular of 16 AdS preons,
corresponding to 1/2-BPS states. From this point of view, it would
be interesting to see whether one can give a non-commutative
counterpart of {\it e.g.} the supermembrane BPS state and, if so,
whether it would be related with the matrix model of a
non-commutative membrane which is used to describe coincident
M2-branes\footnote{For very recent progress in the description of
supersymmetric non-commutative M2-branes see \cite{M2Bagger} and
\cite{Gustavson}.} (see \cite{Berman07} and refs. therein).

\medskip

 {\bf Acknowledgments.} The authors thank Dima
Sorokin for useful discussions. This work has been partially
supported by research grants from the Ministerio de Educaci\'on y
Ciencia (FIS2005-02761), the Ukrainian State Fund for Fundamental
Research (N383), the INTAS (2006-7928), the EU MRTN-CT-2004-005104
'Forces Universe' network and the Spanish Consolider-Ingenio 2010
Programme CPAN (CSD2007-00042).

\end{document}